\documentclass{ceab}   

\usepackage{epsfig}     
\usepackage{graphicx}   

\usepackage{ceabbib}     
\usepackage[T1]{fontenc}
\newcommand{\cnt}{C^\mathrm{nt}}
\newcommand{\cf}{C\!F}

\begin{document}
\title{Far-IR and radio thermal continua in solar flares}
\def\gore{Far-IR and radio thermal continua in solar flares}
\author{J. KA\v{S}PAROV\'{A}$^{1}$, P. HEINZEL$^{1}$, M. KARLICK\'{Y}$^{1}$,  Z. MORAVEC$^{2}$\\
and M. VARADY$^{2,1}$ 
\vspace{2mm}\\
\it $^1$Astronomick\'{y} \'{u}stav Akademie v\v{e}d \v{C}esk\'{e} republiky, v.v.i., Fri\v{c}ova 298, \\ 
\it 251 65 Ond\v{r}ejov, Czech Republic\\
\it $^2$Katedra fyziky, Universita J. E. Purkyn\v e, \v Cesk\'e ml\'ade\v ze 8,\\ 
\it 400 24 \'Ust\'\i~nad Labem, Czech Republic
}
\maketitle
\begin{abstract}
With the invention of new far-infrared (FIR) and radio mm and sub-mm instruments (DESIR on SMESE satellite, ESO-ALMA),
there is a growing interest in observations and analysis of solar flares in this so far unexplored wavelength region. Two principal
radiation mechanisms play a role: the synchrotron emission due to accelerated particle beams moving in the magnetic field and the
thermal emission due to the energy deposit in the lower atmospheric layers. 
In this contribution we explore the time-dependent effects of beams on thermal FIR and radio continua. We show how and
where these continua are formed in the presence of time dependent beam heating and non-thermal excitation/ionisation of the
chromospheric hydrogen plasma. 
\end{abstract}
\keywords{solar flares, radiative hydrodynamics, continuum emission}
\section{Introduction}
New ground-based and space-borne instruments are being designed to observe the Sun in so-far unexplored wavelength windows or 
with unprecedented spatial and/or  temporal resolution. Here we are motivated by the planned observations of solar flares
in far infrared (FIR) region and in sub-millimetre/millimetre (SMM) radio wavelengths. A specific FIR window where the radiation is not
observable from the Earth is between 35 and 250~$\mu$m and this is expected to be explored by the DESIR telescope on board the SMESE
satellite (SMall Explorer for Solar Eruptions -- \citet{2007AdSpR..40.1787V}). At SMM wavelengths, observations were already performed by
\citet{2000ASPC..206..318K} and \citet{2004A&A...415.1123L}. However, new data are expected soon from ALMA 
(Atacama Large Millimeter Array) radio-interferometer.
In both FIR and SMM domains a superposition of two radiation components is expected: the high-frequency part of 
the microwave spectrum from ultra-relativistic electrons and/or positrons (non-thermal synchrotron emission) and the low-frequency
part of the thermal flare continuum. As suggested in \citet{2007AdSpR..40.1787V}, these emissions provide, respectively,
a unique diagnostic of the non-thermal particles (beams) and of the thermal response of the lower flare atmosphere to the energy deposit.
In FIR domain, the relative importance of both processes was estimated by \citet{1975SoPh...43..405O}. Recently, 
\citet{2004A&A...419..747L} and \citet{he08} computed the microwave thermal continua formed in dynamic and static semi-empirical 
atmospheres, respectively.

In this study, we concentrate on formation of such thermal emission in the 35~$\mu$m -- 1~cm wavelength range
originating in  a highly dynamical flare atmosphere heated by short-duration electron beam pulses. 
Section~2 describes the mechanisms responsible for the thermal emission, characteristics of the emission
corresponding to simulations of the beam heated atmosphere are presented in Section~3. Section~4
summarises our results.

\section{Mechanisms of the FIR and SMM thermal continuum formation}\label{sec:thermal_mech}
Depending on the wavelength, various atmospheric depths can contribute to the emergent FIR and SMM intensity \citep{he08}.
In the chromosphere, the dominant source of opacity is the hydrogen free-free continuum. The absorption coefficient
is given as \citep{1979rpa..book.....R}
\begin{equation}
\kappa_\nu(\mathrm{H})= 3.7\times 10^8 T^{-1/2} n_\mathrm{e} n_\mathrm{p} \nu^{-3} g_\mathrm{ff} 
\end{equation}
where $n_\mathrm{e}$ and $n_\mathrm{p}$ are the electron and proton densities, respectively, 
$T$ is the temperature, and $g_\mathrm{ff}\approx 1$ is the Gaunt factor. At lower temperatures, around the temperature minimum region,
H$^{-}$ free-free opacity plays an important role. The H$^{-}$ absorption is \citep{1970SAOSR.309.....K}
\begin{equation}
\kappa_\nu(\mathrm{H}^-)=  \frac{n_\mathrm{e} n_\mathrm{H}}{\nu} (1.3727\times10^{-25} + (4.3748\times10^{-10} - 2.5993\times10^{-7}/T) / \nu)\,
\end{equation}
where $n_\mathrm{H}$ is the neutral hydrogen density. Total absorption corrected for stimulated emission is then
\begin{equation}
\kappa_\nu = \left[\kappa_\nu(\mathrm{H}) + \kappa_\nu(\mathrm{H}^-)\right] (1 - \mathrm{e}^{-h\nu/kT}) \ ,
\end{equation}
where $h$ and $k$ are the Planck and Boltzmann constants, respectively.
The source function for these free-free processes is the Planck function $B_\nu$. However, the absorption has to be evaluated using
number densities coming from non-LTE calculations. Finally, emergent intensity $I_\nu$ can be obtained as
\begin{equation}\label{eq:int}
I_\nu = \int\eta_\nu \mathrm{e}^{-\tau_\nu}\mathrm{d}z\qquad \eta_\nu =  \kappa_\nu B_\nu\quad 
\mathrm{d}\tau_\nu = -\kappa_\nu \mathrm{d}z\ ,
\end{equation}
where $z$ is the geometrical height.

\section{Flare models}\label{sec:models}
We have  evaluated the FIR and SMM continuum emission for non-LTE radiative hydrodynamic models which describe the atmospheric response to the
time dependent electron beam heating. In these simulations electron beams of a varying beam flux and power-law spectra
are injected into a loop which corresponds to an initially hydrostatic VAL C atmosphere \citep{val81}. Then, the time evolution 
of the beam particles and plasma properties are followed. We also consider the so-called non-thermal 
collisional rates, $\cnt$, which account for the changes of the atomic level populations due to the collisions with the beams. 
For more details on the models see \citet{va05a,ka08}.
\begin{figure}[t]
\begin{center}
\includegraphics[width=0.6\textwidth]{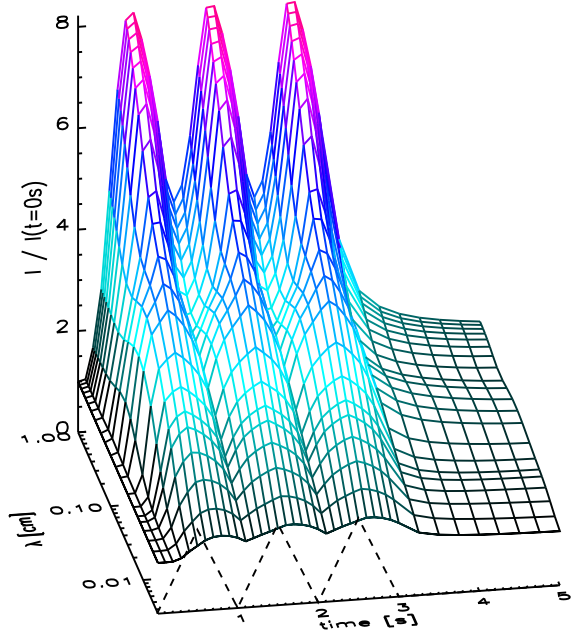}
\caption{Time evolution of relative continuum intensities scaled to the intensity at $t=0$~s, i. e. of the VAL C atmosphere,
for an electron beam of a power-law index $\delta=3$ and a maximum beam flux $F_\mathrm{max} =6\times10^{10}$~erg~cm$^{-2}$~s$^{-1}$. The 
dashed line indicates the beam flux time variation.}
\label{fig:timevar}
\end{center}
\end{figure}
\subsection{Time variation}
Our simulations show that the continuum emission is well correlated with the beam flux 
on the time scale of the beam flux variation, e.g. on a subsecond
time scale. Longer wavelengths (1~cm) exhibit larger relative intensity increase than shorter wavelengths (35~$\mu$m) when compared to the 
the preflare (VAL C) intensities -- see Figure~\ref{fig:timevar}. Furthermore, emission in the $\lambda < 0.2$~mm range depends also on
the beam parameters, i.e. the power-law index and the beam flux. Generally, maxima of the emission lag behind the beam flux maxima, the time lag
lies within the 0.1~--~0.3~s interval for present simulations.
\subsection{Influence of non-thermal collisional rates}
\begin{figure}[t]
\begin{center}
\includegraphics[width=0.65\textwidth]{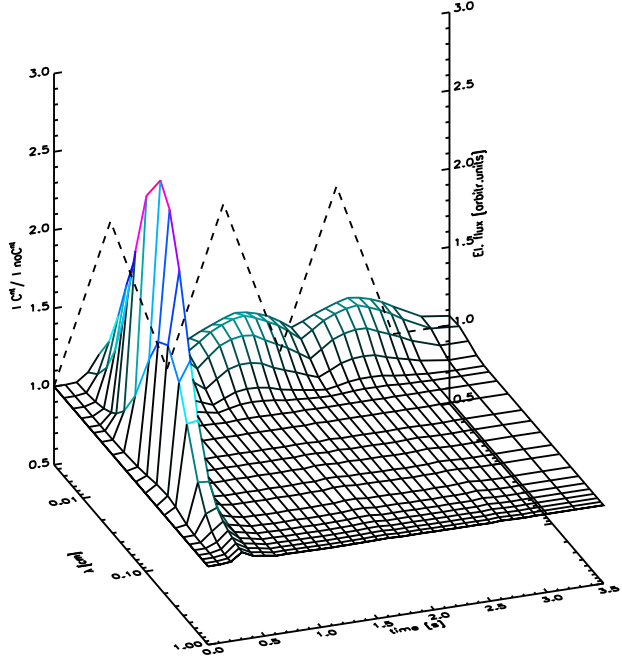}
\end{center}
\caption{Influence of the non-thermal collisional rates $\cnt$. 
The plot shows the ratio of intensities with and without $\cnt$. The dashed line indicates the beam
flux time variation. The beam parameters are the same as in Fig.~\ref{fig:timevar}. Note the reversed wavelength scale with regard to 
Figure~\ref{fig:timevar}.}
\label{fig:cnt_nocnt}
\end{figure}

Although the non-thermal rates can significantly affect the hydrogen level populations and electron densities in some atmospheric layers
\citep{ka05c}, their influence on FIR and SMM continuum emission is almost negligible.
$\cnt$ cause only moderate and temporary increase of the continuum intensity shortly after the beam injection into the VAL C atmosphere.
Additionally, influence of $\cnt$ for $\lambda < 0.2$~mm depends on the beam parameters and it is modulated by the beam flux time variation 
-- see Figure~\ref{fig:cnt_nocnt}.
\subsection{Formation depths}
\begin{figure}[t]
\begin{center}
\includegraphics[width=0.49\textwidth]{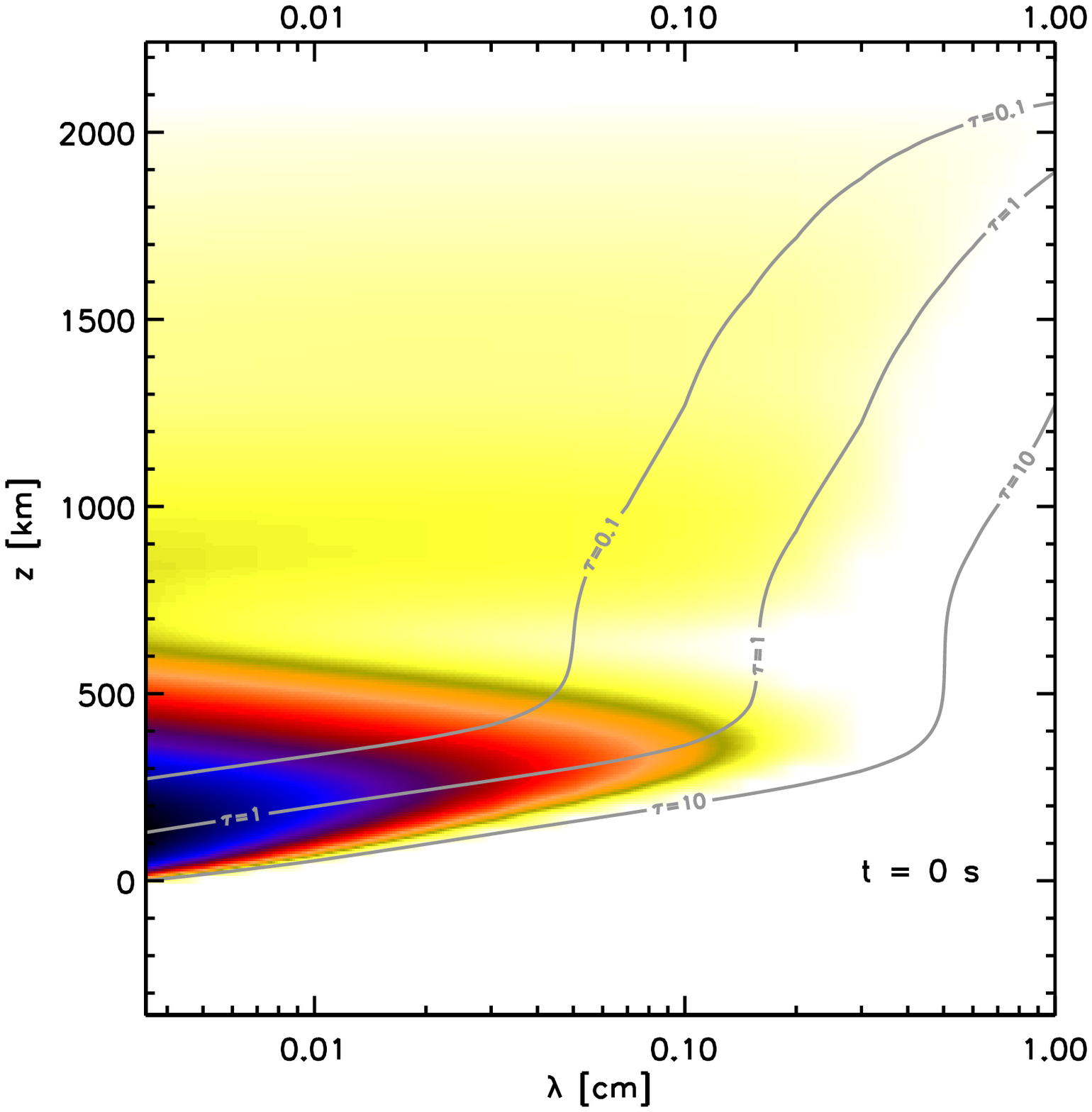}
\includegraphics[width=0.49\textwidth]{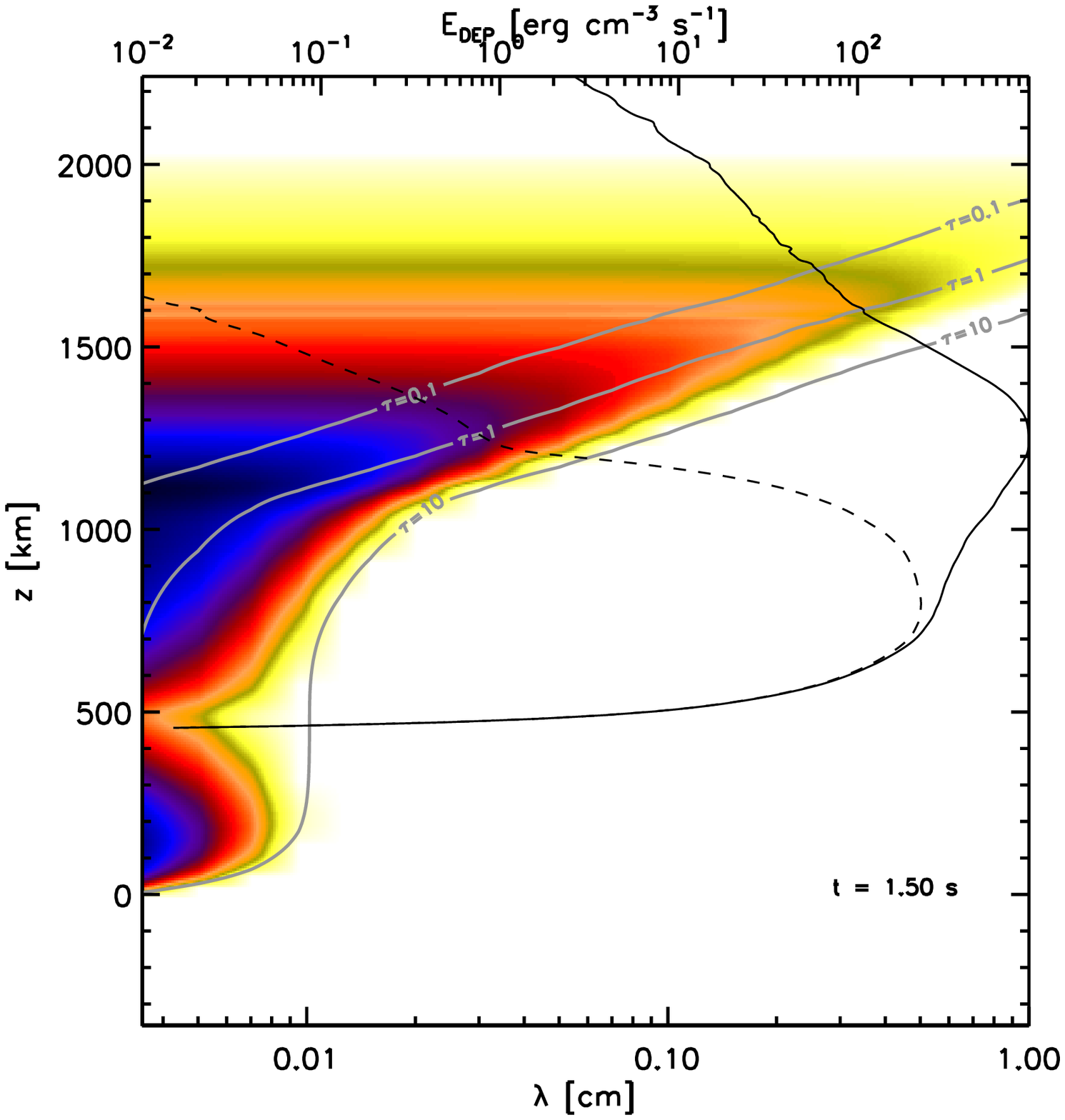}
\end{center}
\caption{An example of the time evolution of contribution function $\cf$ (Eq.~\ref{eq:cf}) for an electron beam of $\delta=3$ 
and $F_\mathrm{max}=4.5\times10^{10}$~erg~cm$^{-2}$~s$^{-1}$. 
{\it Left:} $\cf$ for the VAL C atmosphere, i.e. at $t=0$~s. {\it Right:} $\cf$ at $t=1.5$~s corresponding
to the model with included $\cnt$. 
Colour scale going from white to black represents the range of $\log\cf$ values (from minimum to maximum).
Black solid curve indicates the total energy deposit of the beam, 
dashed curve is for the energy deposit on hydrogen $E_\mathrm{H}$. $\cnt$ are proportional to $E_\mathrm{H}$, therefore the
dashed line marks the layers which are affected by $\cnt$. Gray lines display the levels of optical depth $\tau=0.1,1,10$.}
\label{fig:CF}
\end{figure}

Atmospheric layers which contribute the most to the outgoing intensity, i.e. the formation depths, can be understood
in terms of the so-called contribution function $\cf$
\begin{equation}\label{eq:cf}
\cf = \eta_\nu \mathrm{e}^{-\tau_\nu} \qquad I_\nu = \int\cf\,\mathrm{d}z\, ,
\end{equation}
compare with Eq.~(\ref{eq:int}). In the initial atmosphere, emission at wavelengths shorter than $\lambda\approx 1$~mm
comes from  deep, photospheric layers, whereas the continuum at other wavelengths is formed at the upper part
of the atmosphere. However, as the atmosphere is heated by the beams, the emission is formed mainly at the chromospheric heights
(above $z\approx 1000$~km). The emission which comes from these layers can be influenced also by $\cnt$ since the
beams deposit their energy there. On the other hand, the emission in the 35~--~200~$\mu$m range partially originates from the
photosphere -- see Figure~\ref{fig:CF}.
\section{Conclusions}\label{sec:summary}
Using the results of non-LTE radiative hydrodynamic simulations we have found that the rapidly varying beam flux
can manifest itself in the thermal FIR and SMM continuum intensities. We show that the continuum intensities do vary on the beam flux
variation time scales. However, the direct influence of the beam electrons via the non-thermal collision is moderate only. The emission
is mainly affected by the temperature variations which result from the time dependent heating by the beams.
Concerning the formation of the emission, in the heated atmosphere or due to the non-thermal collisions the continua are
formed in the upper atmosphere. The intensities in the  35~--~200~$\mu$m partially originate also from the photosphere.

In this work we did not consider 
effects related to the situation when the frequency of the emitted electromagnetic waves is close to the local
plasma frequency.
In the case of the solar atmosphere plasma, these effects
must be taken into account at the radio frequency range. We will address this issue for the FIR and SMM emission in a future study.
\section*{Acknowledgements} 
We acknowledge grants 205/04/0358, 205/06/P135, 205/07/1100 (GA CR)
and the research project AV0Z10030501 (Astronomick\'{y} \'{u}stav).
The non-LTE radiative hydrodynamic simulations were performed on OCAS  (Ond\v{r}ejov Cluster for Astrophysical Simulations) 
and Enputron, a computer cluster for extensive computations (Universita J. E. Purkyn\v{e}).
\bibliographystyle{ceab}
\bibliography{kasparova}
\end{document}